\documentstyle[epsfig]{mn}

\newcommand{\gamp}{\gamma_1}
\newcommand{\gamx}{\gamma_2}
\newcommand{\bn}{\hat{\bf n}}

\newcommand{\kp}{{\bf k_{\perp}}}
\newcommand{\kpp}{{\bf k^{\prime}_{\perp}}}

\begin{document}
\onecolumn

\title{Theoretical Estimates of Intrinsic Galaxy Alignment}
\author[Mackey, White \& Kamionkowski]
       {Jonathan Mackey$^1$, Martin White$^1$ and Marc Kamionkowski$^2$\\
$^1$Harvard-Smithsonian Center for Astrophysics, 60 Garden St., Cambridge 
MA 02138\\
$^2$Mail Code 130-33, California Institute of Technology, Pasadena, CA 91125}
\date{June, 2001; in final form May, 2002}
\maketitle

\begin{abstract}
It has recently been argued that the observed ellipticities of
galaxies may be determined at least in part by the primordial tidal
gravitational field in which the galaxy formed.  Long-range
correlations in the tidal field could thus lead to an
ellipticity-ellipticity correlation for widely separated galaxies.  We
present a new model relating ellipticity to angular momentum, which
can be calculated in linear theory. We use this model to calculate the
angular power spectrum of intrinsic galaxy shape correlations. We show
that for low redshift galaxy surveys, our model predicts that
intrinsic correlations will dominate correlations induced by weak
lensing, in good agreement with previous theoretical work and
observations.  We find that our model produces `$E$-mode' correlations
enhanced by a factor of $3.5$ over $B$-modes on small scales, making
it harder to disentangle intrinsic correlations from those induced by
weak gravitational lensing.
\end{abstract}

\begin{keywords}
gravitational lensing - cosmology: theory - large-scale structure
of Universe
\end{keywords}

\section{Introduction}
Galaxy alignments, or ellipticity correlations, have been the subject
of much study over many years (Djorgovski~\shortcite{Djo87}, and
references therein), extending back well into the 1800s.  Once
galaxies were discovered and were found to be elliptical in shape, the
logical next step was to see if they have a tendency to align with
each other.  This would provide a clue as to what they were and how
they formed.  More recently it was realised that different models of
structure formation may produce different levels of alignment, and so
searches aimed at placing constraints on these models.
Searches have historically concentrated in nearby regions
of high galaxy density (clusters and superclusters) in order to get
good number statistics.  The results have been, in general, inconclusive or
contradictory of previous results, although weak trends for
alignments have been found for some samples \cite{CabAld98}.  

Recently, there has been renewed theoretical interest in long-range
correlations of galaxy ellipticities.  While initially the
interest was in reconstructing the gravitational potential field from
the orientations of galaxy spins \cite{LeePen00a}, it has been driven
lately by the increasing success of field-surveys for weak
gravitational lensing by large scale structure (see
Mellier~\shortcite{mellier99} for a recent review).  Weak lensing
shear, the coherent distortion of galaxy images on the sky induced by
density perturbations along the line of sight
\cite{gun67,mir91,bla91,kai92,bar92,bar99} has now been detected by
several different groups
\cite{Waeetal00,BacRefEll00,Witetal00,KaiWilLup00,Maoetal01,RhoRefGro01}.
Following standard practice, all of these authors assume that all of their
observed correlations in the ellipticities of galaxies comes from weak lensing,
with an immeasurably small intrinsic signal.
Intrinsic correlations, if present, may contaminate the weak lensing signal
and so they should be considered when interpreting results from these
field-lensing surveys.

Theoretically, alignments are expected at some level because nearby galaxies
form in similar and related gravitational fields; they should therefore react
in similar ways to the influence of this field and so should have some tendency
to align with each other.
Several authors have revisited this question recently, with simulations
\cite{HeaRefHey00,CroMet00} and analytic arguments
\cite{LeePen00a,CatKamBla00,cnpt00a}.
These authors assume that the ellipticities of the observed luminous galaxies
are determined by either the shapes of their dark matter halos or by the halo
angular momenta.  In either case, the ellipticity should ultimately be
determined, at least in part, by the tidal gravitational field in which the
galaxy forms.  The existence of long-range correlations in the tidal 
field~\cite{CatPor01} 
will thus lead to correlations in ellipticities of widely separated sources.
Observationally, Pen, Lee \& Seljak~\shortcite{penleesel00} report a tentative
detection of galaxy alignments in the local universe using the Tully catalog.
More recently, Brown et al.~\shortcite{brownetal00} find a more significant
correlation using data from the SuperCOSMOS Sky 
Survey\footnote{http://www-wfau.roe.ac.uk/sss/}.
These detections reinforce the importance of considering intrinsic alignments.

An analytic theory of these correlations, even if only approximate, can provide
valuable guidance on the plausible shape, size, redshift evolution and sample
depth dependence of the effect as well as flag potential indicators of
contamination.  
Catelan et al.~\shortcite{CatKamBla00} (hereafter referred to as CKB) 
calculated ellipticity correlations under the assumption that a galaxy's 
ellipticity is determined by the initial halo shape.  
They also outlined an argument for how a galaxy's angular momentum may 
determine its ellipticity in the context of the tidal torque theory using 
the Zel'dovich approximation.
A similar model was developed more fully by Crittenden et
al.~\shortcite{cnpt00a} (hereafter referred to as CNPT) who calculated
correlations in ellipticity due to correlations in the direction of the
galaxy's angular momentum vectors.
Their model predicts that for low redshift galaxy surveys intrinsic alignments
are expected to dominate over weak lensing correlations, with considerable
uncertainty in the normalisation.  These predictions were later confirmed 
 by the measurements of Brown et al.~\shortcite{brownetal00}.
In a later paper \cite{cnpt00b} they also
discuss ways of discriminating between intrinsic correlations and weak lensing
induced correlations.

In this paper we build upon some of the ideas presented in CKB, developing an
alternate and complementary model to that of CNPT.  We calculate in
our model how correlations in ellipticity, or ``intrinsic shear,'' can
be generated by correlations in angular momentum, with an emphasis on
seeing how large intrinsic correlations can possibly be.  While some of the
input physics of our model is the same as that of CNPT, our approach
and analysis are rather different.  CNPT calculate correlations in the
direction of the angular momentum only, and for the case where the
inertia tensor and tidal field tensor of the protogalaxy are strongly
correlated.  Our model assumes that these tensors are uncorrelated 
(numerical simulations have recently produced evidence that this assumption
is not justified, see \S\ref{sec:ttt}, although we argue
that this assumption does not affect our results significantly),
which results in a significantly different expression for the angular
momentum generated. 
It also includes correlations in the magnitude of
ellipticities, which may be present.  The two models are 
normalised in different ways, to different physical quantities.  In
our calculation, we have tried to make approximations that will, if
anything, overestimate the ellipticity correlations, in order to place
upper limits on their magnitude.

All of these models are really educated guesses as to the relationship between
ellipticity and angular momentum, and are also chosen at least partly because
of mathematical simplicity.
Ideally we would like a quantitative model of galaxy formation which could tell
us the relationship between galaxy shape/alignment and the environment in which
a galaxy forms.  We are currently far from this ideal situation, and so must
make certain assumptions for our model which we believe are physically well
motivated.
It is thus both surprising and gratifying that we obtain very comparable
results to CNPT.  This may indicate that both our and their conclusions are
robust to the considerable simplifications which need to be made in such
calculations.

The outline of this paper is as follows.
In the next section we discuss our model for relating the angular momentum
of a galaxy to its ellipticity in the context of tidal torque theory.
In \S\ref{sec:calculation} we present our main calculation, explaining in
\S\ref{sec:normalization} how we normalise our model and showing in
\S\ref{sec:resultsummary} how to relate our results to other measures of
correlations.  The reader not interested in the details may find the
relevant equations summarized in \S\ref{sec:resultsummary}.
We present results for an example ($\Lambda$CDM) model in \S\ref{sec:results},
explain them in terms of the underlying physics, and consider the effects of
intrinsic correlations  on weak lensing measurements of `cosmic shear.'
Then in \S\ref{sec:comparison}, we discuss our results in the context of 
observations and previous work.
We present our conclusions in \S\ref{sec:conclusions}.

\section{Background} \label{sec:background}
In this section we discuss the ingredients of our model.  First we
introduce the concept and mathematical description of the ellipticity of
galaxies.  The key assumption we make is that the ellipticity
(orientation and magnitude) is determined by the galaxy's angular
momentum.  We fix our model so that the ellipticity components
transform correctly under rotations in the plane of the sky, and so
that the magnitude of the ellipticity of galaxies with more angular
momentum is larger.  We implicitly assume that the angular momentum of
the luminous galaxy is aligned with that of the dark matter halo it
resides in.  The angular momentum of a galaxy can be predicted and
related to the local gravitational potential (in linear theory) using
`tidal torque theory'.  This then allows us to calculate correlations
in the ellipticity in terms of correlations in the gravitational
potential, which is well specified in linear theory for a given
cosmological model.

\subsection{Galaxy ellipticities}

It is a standard assumption in calculations of weak gravitational lensing that
the source galaxies are randomly oriented with respect to each other.  We 
wish to investigate to what extent this holds true.
Lensing will introduce a shear in the shapes of the sources, so we need to
describe any ``intrinsic'' shear that these sources have, which we do in terms
of the spin-2 (complex) ellipticity:
\begin{equation}
  {\bf \epsilon} = |\epsilon|e^{2i\phi}  = \gamma_{1}+i\gamma_{2} \;.
\end{equation}
The components $\gamma_{i}$ are usually defined in terms of the 
second moments of the light distribution of a galaxy, $F(x,y)$, as
\begin{eqnarray}
  \gamp &=& { \int F(x,y)(x^2-y^2)dxdy \over \int F(x,y)(x^2+y^2)dxdy} \;,
\nonumber \\
  \gamx &=& { \int F(x,y)(2xy)dxdy \over \int F(x,y)(x^2+y^2)dxdy} \;.
\end{eqnarray}
It is clear that their values for a given galaxy will depend on the 
orientation on the sky of the coordinate axes chosen.
The correlation functions $\langle \gamma_{i} \gamma_{j}' \rangle$ are 
obtained by measuring $\gamma_{i}$ at one galaxy and $\gamma_{j}'$ at 
another galaxy some distance away on the sky and then multiplying them, 
doing this for many galaxies separated by the same angular distance, and
averaging the result to get the correlation function.
Intrinsic shear correlations can be due to correlations in both the
magnitude and the orientation of the ellipticity across the sky, both
of which are encoded in the components $(\gamp,\gamx)$.  For
example, if all galaxies were flattened disks with axes of symmetry
perfectly parallel with each other, then the correlation would be
unity on all scales.  If the disks have finite thickness, however, even if 
galaxies are perfectly aligned, variations in $|\epsilon|$ with position will 
cause the correlation to vary with separation. 
In the opposite limit, if all galaxies are randomly oriented and have 
random $|\epsilon|$, then the correlations are zero and there is no alignment.

We assume that this ellipticity is determined, at least in part, by
the galaxy's angular momentum (and hence by the initial tidal field as
we show below).  For spiral galaxies this is well justified, as the
disk angular momentum is perpendicular to the plane of the disk.
There is some question as to how closely the angular momentum of the
gas and stars in the disk follows that of the dark matter \cite{AbeCroHer01}, 
but it seems reasonable that they should be closely related.
Elliptical galaxies tend not to have much net angular momentum, and
the ellipticity seems to be determined largely by the velocity
dispersion along the principal axes of the 3D ellipsoid.  Thus
our calculation may be less relevant for elliptical galaxies than
for spirals, and the halo shapes calculation of CKB may be more 
applicable.

Following the angular momentum--ellipticity discussion in CKB, 
using the assumption that the disk of a galaxy forms 
perpendicular to the angular momentum, the observed ellipticity of a galaxy 
will be 
\begin{eqnarray}
\gamp &=& f(L,L_{z}) (L_{x}^{2}-L_{y}^{2}) \,,
\nonumber \\
\gamx &=& 2 f(L,L_{z}) L_{x}L_{y} \,,
\label{eqn:ellipdef}
\end{eqnarray}
where we take the sky to be the {\em x-y} plane, and where
$f(L,L_{z})$ is an unknown function which determines how ellipticity
scales with $L$.  Since we have no firm theory to predict the form of
$f(L,L_{z})$ we will take it to be a constant, $C$, whose value must
be fitted empirically to the observed rms ellipticity of galaxies.
As a result, in our model the ellipticity of a galaxy scales
quadratically with its angular momentum, so galaxies with more angular
momentum will be more flattened.  For the case where galaxies are
approximated as thin disks, it can be shown that
$f(L,L_{z}) = 1/(L^2 + L_{z}^2)$ (see CNPT), thus taking out the dependence
on $L$.  In this case $|\epsilon|$ is set to a constant independent of $L$ and
only the orientation is measured by $\gamma_{i}$.

The price we pay for letting $|\epsilon|$ vary with $L$ is that galaxies in
the tail of the distribution with $L \gg \langle L \rangle$ can potentially
have an unphysical $|\epsilon| >1$.
We have calculated the fraction of galaxies this will apply to.  
In our model ellipticity is proportional to the square of the the
gravitational potential (as we show in the next section), and so if we
assume the potential fluctuations are gaussian distributed, we can
calculate the ellipticity probability distribution from this.  
The rms ellipticity of galaxies (which we take to be
$\bar{\epsilon} = 0.4$; see \S\ref{sec:normalization}) gives the
variance of $|\epsilon|$.  We can thus integrate over the ellipticity
probabilitiy distribution to find what fraction of galaxies have
$|\epsilon| >1$, and we find that this is $\sim 4 \%$.  Larger
values of $|\epsilon|$ are exponentially rare, so they will not affect
our calculation in a significant way.

\subsection{Tidal torque theory} \label{sec:ttt}

Hoyle~\shortcite{Hoy49} was the first to suggest that galaxies acquire their
angular momentum by tidal torques due to the surrounding matter
distribution acting on the protogalaxy.  This torque comes about from
a misalignment of the protogalaxy's mass distribution (inertia tensor)
with the local tidal field.  Using the Zel'dovich approximation
Doroshkevich~\shortcite{dorosh70}, and subsequently White~\shortcite{white84},
showed that the angular momentum acquired by a protogalaxy to first
order in the gravitational potential is
\begin{equation}
  L_{i}(t) \propto \epsilon_{ijk} T_{jl} I_{kl},
\label{eqn:angmom}
\end{equation}
where $T_{ij}$ is the tidal field tensor at the galaxy's centre of mass
and $I_{ij}$ is the inertia tensor of the protogalaxy.

The inertia tensor is given by
\begin{equation}
I_{ij}=\int_{\Gamma} (q_{i}-\bar{q}_{i})(q_{j}-\bar{q}_{j})
   \rho_0a^3\ d^3{\mathbf q} \,,
\end{equation}
where $\Gamma$ is the Lagrangian volume of the protogalaxy, 
${\mathbf q}$ is the Lagrangian position vector of the constituent particles 
with centre of mass ${\mathbf \bar{q}}$, and $\rho_0$ is the
background density.
We follow CKB in assuming each galaxy has the same eigenframe
moment of inertia and the eigenframes are distributed isotropically.
Further we suppose that two of the three principle moments are equal, so that
the unequal moment defines a symmetry axis
$\bn = (\cos \theta, \sin \theta \cos \phi, \sin \theta \sin \phi)$,
and so the inertia tensor simplifies to $I_{ij} \propto n_{i}n_{j}$.

The tidal field tensor is given by
\begin{equation}
  T_{ij} \propto \partial_{i}\partial_{j}\Phi({\bf x}) \,,
\end{equation}
where $\Phi$ is the gravitational potential.  All of the
proportionality above is time-dependent, but we assume that all our
galaxies form at a similar redshift, and so absorb all time dependent
quantities into the constant $C$ above.  
We normalise our theory in such a way that any time dependence will cancel 
out.

By inspection of equation (\ref{eqn:angmom}) it can be seen that if
${\bf I}$ and ${\bf T}$ have the same principal axes then there is no angular
momentum acquired to first order.  There must be some misalignment
between them to generate angular momentum, so the tensors cannot be
perfectly correlated.  In our calculation we make the simplest
assumption, namely that they are completely uncorrelated.  Reality
lies somewhere in between the extremes of perfectly correlated and
uncorrelated.  Nevertheless, our expectation is that the large
wavelength Fourier modes that are primarily responsible for the long
range {\em correlations} in the tidal field should be statistically
independent of the smaller wavelength Fourier modes that are primarily
responsible for the inertia tensor.  So the hope is that our results 
should not be very sensitive to this assumption.
This assumption comes into play when we calculate the ellipticity 
correlation function between neighbouring galaxies.  We must average
over the distribution of inertia tensors and tidal tensors.  If the two
are uncorrelated we can average over inertia tensors first (as we expect
the ellipticity correlation to be due to long range tidal field 
correlations), and subsequently over the tidal field.  Because this 
assumption is an important part of our calculation, we discuss relevant
numerical results in the next section.

\subsection{Tidal Torque Theory in Numerical Simulations} \label{sec:TTT}

\subsubsection{Initial Angular Momentum}
Several authors have investigated the questions of angular
momentum generation and tidal torque theory (TTT) in numerical simulations.
Sugerman, Summers \& Kamionkowski~\shortcite{SugSumKam00} found a relatively good
correlation between the predictions of linear theory and the actual
evolution of dark matter halos.  They found that the Zel'dovich
approximation we have used typically overpredicts the magnitude of the
angular momentum generated by a factor of $\sim 3$, and that the spin
direction is marginally reproduced with a large scatter.

Lee \& Pen~\shortcite{LeePen00a} differed from Sugerman et
al.~\shortcite{SugSumKam00} in that they found quite a strong
correlation between the predicted and simulated spin direction, again
with a large scatter.  They concluded that TTT is a reasonably good
predictor of the direction of the angular momentum vector of a halo.

Porciani, Dekel \& Hoffman~\shortcite{PorDekHof01a}, in a more
detailed study of TTT, used dark matter simulations to look at the
spin (i.e.\ angular momentum) amplitude and direction of halos, as
well as spin--spin correlations between halos.  They reproduce the
result of Sugerman et al.~\shortcite{SugSumKam00} in finding that TTT
systematically overpredicts that amplitude of the spin, but by a
constant fraction on average, with a large scatter.  The agreement
between these two groups is reassuring, and gives us confidence in
our calculation. 
A constant bias in the magnitude of ${\bf L}$ will be taken out by the
normalisation of our model, so that the `angular momentum' we
calculate really does correspond (on average) to the spin amplitude of
a halo.

As regards spin direction, they find that at high redshift, TTT 
predictions are in
good agreement with their simulations, although there is some
scatter.  They also studied spin--spin correlations of dark matter
halos (i.e. correlations in the directions of the spin vectors among
neighbouring halos) and find that at high redshift the tidal torque
theory reproduces the correlations quite well, albeit with large
scatter.
These results are important for our calculation in that they show 
the theory on which it is based to be reasonably accurate, at least at
high redshift. The situation is not so good at low redshift, which we
return to below.

\subsubsection{Inertia and Tidal Tensor Correlations}
Another important effect is a correlation between the inertia and tidal field
tensors.
Lee \& Pen~\shortcite{LeePen00a} found that ${\bf I}$
and ${\bf T}$ were quite strongly correlated in their simulations (although
misaligned to a detectable degree), and this conclusion was supported
very recently by Porciani, Dekel \& Hoffman~\shortcite{PorDekHof01b}.
They find that angular momentum is generated by the small but significant 
misalignment of the tensors.
One way of thinking about this is to decompose the inertia tensor into
a part perfectly correlated with the tidal tensor, and another part
completely random: ${\bf I}_{\rm tot} = {\bf I}_{\rm corr} +{\bf I}_{\rm random}$ (where
we assume for now that different components of ${\bf I}$ and ${\bf T}$ are all
correlated at the same level).  The correlated part generates no
angular momentum to first order so the angular momentum generated is
just some fraction of what would have been generated if ${\bf I}$ was all
random.  The direction is unchanged, only the magnitude is suppressed.
Such a situation does not affect our result, because the overall
normalization of ${\bf L}$ factors out, as we show in
\S\ref{sec:normalization}.  Catelan \& Theuns~\shortcite{CatThe96}
also argued analytically that assuming ${\bf I}$ and ${\bf T}$ are uncorrelated
will result in an over-prediction of the magnitude of the angular
momentum generated, and this has been verified in simulations
\cite{SugSumKam00,PorDekHof01a}.

We would see an effect due to ${\bf I}$--${\bf T}$ correlation if the direction of
${\mathbf L}$ changed when these two tensors are more strongly
correlated.  One way to make this happen is if higher order terms
become important when this leading order term given by
equation~(\ref{eqn:angmom}) is suppressed.  
Another way is if different components of ${\bf I}$ are correlated with
differing strengths to different components of ${\bf T}$. 
Such a situation could arise if
the first principal axes of ${\bf I}$ and ${\bf T}$ were strongly correlated, but
the second and third were distributed more randomly in the
plane perpendicular to the first.  
Porciani et al.~\shortcite{PorDekHof01b} investigated the correlation
between the different principal axis directions of ${\bf I}$ and ${\bf T}$.  They
found that each principal axis of ${\bf T}$ is strongly correlated with the
corresponding axis of ${\bf I}$, but all at roughly the same level (to
within $\sim 15\%$), indicating that the strength of the correlation is
comparable between the various components, although not identical.
They also find that the initial direction of ${\mathbf L}$ 
(with respect to the local tidal field) for halos in their simulation 
is somewhat
different to the direction they obtain after making ${\bf I}$ and ${\bf T}$
independent by randomizing the orientation of ${\bf I}$, in that the same
trends are seen but they are weaker in the real simulation.  
This could be because of the differing strengths of ${\bf I}$--${\bf T}$
correlations among different components, or it may be because higher
order terms have become more important relative to the suppressed
leading order term, adding larger scatter to the direction of 
${\mathbf L}$ when compared to the ${\bf I}$--${\bf T}$ uncorrelated case.

So where does this leave our calculation?  We expect that the larger
scatter and weaker correlation between ${\mathbf L}$ and ${\bf T}$, as
compared to the predictions when there is no ${\bf I}$--${\bf T}$ correlation,
will cause galaxy alignments to be somewhat weaker than our prediction
by some factor, making our prediction an upper limit.  This is still a
useful constraint, as the CNPT analysis explicity assumes strong
${\bf I}$--${\bf T}$ correlations (although see below) and so cannot be applied 
to the uncorrelated case to give an upper limit.  
Also this simple rescaling should be the
only difference between our results due to the ${\bf I}$--${\bf T}$ correlation,
and so other differences will be due to other differences in our respective
models.  It is important to assess which results are robust and which
are sensitive to the simplifications in both models used to calculate
ellipticity.  We discuss this in detail in \S\ref{sec:comparison}.
The differences seen by Porciani et al.~\shortcite{PorDekHof01b} are 
not large, however, especially when compared to the effects of late-time
non-linear effects also seen in their simulations, so we believe that
${\bf I}$--${\bf T}$ correlations are not affecting the relevance our calculation very
significantly.

\subsubsection{Late--Time Effects on Initial Conditions}
Another issue is the time dependence of a galaxy's angular momentum.
The calculation presented below is a linear theory calculation of the angular
momentum acquired in the initial stages of formation.
We implicitly assume that this subsequent non-linear
evolution of the galaxy and its interaction with its environment do
not substantially alter its angular momentum.  This is almost
certainly false in dense environments such as clusters of galaxies and
compact groups, where the dynamical time is much shorter than the
Hubble time.  We expect, therefore, that our analysis will not apply
to them, but rather only to more isolated galaxies.  The criterion is
that they should not have exchanged significant angular momentum with
neighbouring galaxies/halos over a Hubble time.  This is hard to quantify in
practice given that galaxies form from mergers of smaller objects, so
in a sense all galaxies have had many significant dynamical
interactions in the past.  Further, galaxies are not isolated points
in the universe --- they live in extended halos and it is hard to say
where the halo of one galaxy stops and where the neighbouring ones
start.  

Sugerman et al.~\shortcite{SugSumKam00} and 
Porciani et al.~\shortcite{PorDekHof01a} indeed found that the spin
direction of a dark matter halo changes with time, and that
correlations between the spin direction and the initial conditions
weaken significantly between the good agreement at $z\sim 50$, and
$z=0$.  At $z=0$ there is typically a $30$--$40^{\rm o}$ 
misalignment between the
initial and $z=0$ spin direction.  The resulting distribution is not
random by any means and clearly retains some memory of the initial
conditions, but the spin direction does change significantly, with
most of the change happening between $z=3$ and $z=0$.  The level of
the spin--spin correlation at $z=1$ is of order $1\%$ at a halo
separation of $1h^{-1}$Mpc, which is in good agreement with work by
Croft \& Metzler~\cite{CroMet00} and Heavens et al.~\cite{HeaRefHey00},
who looked for intrinsic halo alignments in large N-body simulations.

The striking feature of the Porciani et al.~\shortcite{PorDekHof01a} 
results is the decrease in the spin--spin correlation by a factor of
$5$ or so, as compared to the correlation at $z=50$.  
The spin--spin correlation at $z=50$ is in line with TTT
predictions, and it is really non-linear effects such as mergers and
angular momentum exchange between halos which reduce it over time.
Lee \& Pen~\shortcite{LeePen00a} introduced a parameter `$a$' to account 
for the effects of ${\bf I}$-${\bf T}$ correlations and non-linear processes 
on the (linear theory) correlation between spin direction and the 
tidal field.  This results in a more random spin direction, and lower
spin--spin correlations by a factor of $a^2$, as compared to the
assumption of independent ${\bf I}$ and ${\bf T}$ and no non-linear effects.  
The Porciani et al. results indicate that it is non-linear 
processes acting cumulatively over time to reduce spin--spin 
correlations which are providing the dominant contribution to 
Lee \& Pen's parameter `$a$'.
This strengthens our argument that ${\bf I}$-${\bf T}$ correlations do not 
significantly affect or invalidate our calculation, and implies that 
we could apply a simple correction factor to the final spin--spin 
correlation result to account for these non-linear effects.
We note that it is not clear from the simulations if this
suppression of spin--spin correlations is scale dependent or not; one
might guess it is not because it seems to result from perturbing spin
directions of galaxies in what is probably a fairly random manner.

Also relevant is a potentially more radical idea proposed by
Vitvitska et al.~\shortcite{Vitetal01}, who investigate the
possibility that angular momentum is generated primarily by mergers
and not by tidal torques.  This is perhaps not unexpected in the
context of hierarchical models of structure formation where galaxies
form by assembly of smaller parts through mergers.  In this situation,
however, the TTT can still give reasonable results even though the
merger details are missed.  When the tidal field is smoothed over a
galaxy-mass region the smaller scale fluctuations that give rise to
the galaxy progenitors are erased, but the sum of all their angular
momenta (which eventually gets into the galaxy through mergers) is
what is calculated by the theory.

\subsubsection{Summary of TTT}
There seems to be some consensus that while tidal torque theory is not perfect, it
works surprisingly well and shows that at least the outlines of what we
are observing are inherent in linear theory.
That TTT works at all is somewhat surprising, considering it
was first thought of in the context of top-down structure formation models
wherein galaxies form via the monolithic collapse of one large concentration
of matter.  
For our purposes, it seems that TTT is a good starting point.
To extend the calculation requires including non-linear effects which will
add `noise' to our predictions.
These non-linear effects will overwrite initial correlations to some extent, 
and it is not yet clear whether they just add a random component to 
the angular momentum or if on small scales they create their
own strong alignments, or both.
Indeed observations by Brown et al.~\shortcite{brownetal00} indicate that
galaxies have a fairly strong alignment in the local universe,
probably significantly stronger than that seen between halos in
simulations (at least on some scales).  It is difficult to compare the
strength, however, because of differing statistics calculated to
measure correlation.

It seems that neglecting ${\bf I}$--${\bf T}$ correlations affects our results
somewhat, but probably not very significantly, and it is not clear
that there is a good analytical way of taking this into account.
More significant is the dilution of spin correlations over time which
is seen in simulations.  This could be dealt with by simply rescaling
the analytically calculated correlations.  It is possible (and perhaps
likely), however, that since galaxies collapse to the very centres of
dark matter halos, their spin direction is much less affected by late
time effects than the outer regions of halos (where most of the
angular momentum is), simply because there is much less of a lever arm
to torque a small compact galaxy.

\section{Calculation} \label{sec:calculation}

In this section we present our main calculation.  The derivation is somewhat
technical and the reader interested in only the main results can skip to
\S\ref{sec:resultsummary}.  Throughout we shall perform our calculation in
Fourier space and express our results in terms of angular power spectra.
This allows us to take advantage of much of the theory of spin-2 fields
developed for the study of polarization of e-m radiation and gravitational
waves \cite{ZalSel97,KamKosSte97,HuWhi97}.  In \S\ref{sec:resultsummary} we 
discuss how to convert our results to real space correlation functions to 
compare with earlier work.

\subsection{Power spectra} \label{sec:powerspectra}

We begin by calculating the power spectra of the ellipticities, which
are each quadratic in the angular momenta (see
equation~(\ref{eqn:ellipdef})).  In outline, we first average our
expressions for the ellipticities of a galaxy over all realisations of
the inertia tensor to obtain an expression in terms only of
derivatives of the gravitational potential.  We then decompose the
spin-2 ellipticity field into scalar (gradient or electric-type modes)
and pseudo-scalar (curl or magnetic-type modes) fields in Fourier space
for ${\bf k}$-vectors perpendicular to the line of sight.  This enables us
to construct 3D power spectra for the ellipticities, which are
convolutions over the density power spectrum.  To proceed from here,
we use the Limber approximation in Fourier space to evaluate the
predicted angular power spectrum for different source distributions.
It is important bear in mind that in the Limber approximation {\it
only\/} modes transverse to the line of sight contribute.  Evaluating
correlation functions in real space is a relatively simple integral
over the power spectra, so we calculate these for comparison with
previous theoretical estimates and observations.

Ignoring the normalization factors for now (\S\ref{sec:normalization}),
we have 
\begin{eqnarray}
  \gamp &=&  L_{x}^{2}-L_{y}^{2} \;,
  \nonumber \\
  \gamx &=&  2 L_{x}L_{y} \;,
  \nonumber \\
  L_{i} &=& n_{l} n_{j} \epsilon_{ijk} \Phi_{,kl} \,.
\end{eqnarray}

As discussed above, we assume that the inertia tensor (i.e. the
direction $\bn$) is independent of the tidal field.  This means we can
average over orientations of the inertia tensor separately to
averaging over realisations of the tidal field.  When we do this, we
get the expectation value for a given tidal field
\begin{equation}
\langle L_{i}L_{j} \rangle =
  {1 \over 15} \left\{ \epsilon_{ikl} \epsilon_{jmn}\Phi_{,lm} \Phi_{,kn}
          - \Phi_{,ik} \Phi_{,jk} + \delta_{ij} \Phi_{,kl}\Phi_{,kl} \right\} \,.
\end{equation}
This equation is the same as equation~(A2) in Lee \& Pen~\shortcite{LeePen00b}, up to an overall normalisation constant.
Using this, we find
\begin{eqnarray}
 \gamp &=& {1 \over 15} \{ (2\Phi_{,zz} - \Phi_{,xx}-\Phi_{,yy})
(\Phi_{,xx}-\Phi_{,yy}) + 3(\Phi_{yz}^{2}-\Phi_{xz}^{2}) \} \,,
\nonumber \\
 \gamx &=& {1 \over 15} \{ 2(2\Phi_{,zz} - \Phi_{,xx}-\Phi_{,yy})\Phi_{,xy}
                        -6\Phi_{,xz}\Phi_{,yz}  \} \,,
\end{eqnarray}
where now $\gamma_{i}$ have been averaged over the distribution of inertia tensors.
The projection of these ellipticities onto the plane of the sky will
be a spin-2 field as in weak lensing analysis.  We can construct
scalar (grad or $E$-mode) and pseudo-scalar (curl or $B$-mode) functions 
of the ellipticities
by taking derivatives.  We do this in Fourier space, because the
analysis is somewhat simpler and more transparent, and in the end
provides a clearer physical understanding of our results.  Because the
ellipticity is quadratic in the gravitational potential, the Fourier
transform 
$\widetilde{\gamp}({\mathbf k}) \equiv \int d^3 {\mathbf x} \gamp({\mathbf x}) 
e^{-i{\mathbf k \cdot x}}$  
is a convolution over the
Fourier modes of the potential.  The $E$- and $B$-modes are given by 
\cite{Ste96,Kametal98}
\begin{eqnarray}
 \epsilon({\mathbf k})k^2 &=&
  (k_x^2-k_y^2) \widetilde{\gamp}({\mathbf k})
     +2k_xk_y\widetilde{\gamx}({\mathbf k}) \,,
\nonumber \\
 \beta({\mathbf k})k^2 &=&
   -2k_xk_y \widetilde{\gamp}({\mathbf k})
       + (k_x^2-k_y^2) \widetilde{\gamx}({\mathbf k}) \,.
\end{eqnarray}
Note that we will we working entirely in the flat sky approximation, 
as any correlations present will go to zero on large scales.
So we obtain
\begin{eqnarray}
 \epsilon({\mathbf k})k^{2} =
 {1\over 15}\int {d^3{\mathbf k'}\over (2 \pi)^3}
        \tilde{\Phi}({\mathbf k'})\tilde{\Phi}({\mathbf k-k'}) 
        f_{\epsilon}(\kpp,\kp -\kpp,k_{z}') \,,
\nonumber \\
 \beta({\mathbf k})k^{2} = 
 {1\over 15}\int {d^3{\mathbf k'}\over(2 \pi)^3}
        \tilde{\Phi}({\mathbf k'})\tilde{\Phi}({\mathbf k-k'}) 
        f_{\beta}(\kpp,\kp -\kpp,k_{z}') \,,
\label{eqn:ebmodes}
\end{eqnarray}
where ${\bf k'}$ is the convolution variable, and $f_{\epsilon,\beta}$ are
functions containing all the information on the relative orientations
of ${\bf k}$ and ${\bf k'}$ and hence the information on the
derivatives of $\Phi$.  They are given by
\begin{eqnarray} 
f_{\epsilon}({\mathbf a,b,} c) &=&
  {1\over 2} (2c^2-a^2)(b^4+{\mathbf (a\cdot b)^2-(a \times b)}^2
  +2b^2({\mathbf a \cdot b}))
\nonumber \\
& &+ {1\over 2} (2c^2-b^2)(a^4+{\mathbf (a\cdot b)^2-
  (a \times b)}^2+2a^2({\mathbf a \cdot b}))
\nonumber \\
& &+ 3 c^2({\mathbf (a+b)^2(a \cdot b)}+2{\mathbf(a \times b)}^2) \,,
\nonumber \\
\nonumber \\
f_{\beta}({\mathbf a,b},c) &=& (2c^2-a^2)(b^2+{\mathbf (a\cdot b)})
{\mathbf (a \times b)}
\nonumber \\
& &- (2c^2-b^2)(a^2+{\mathbf (a\cdot b)}){\mathbf (a \times b)}
\nonumber \\
& &+ 3c^2(a^2-b^2)){\mathbf (a \times b)} \,.
\label{eqn:ebmodefunctions}
\end{eqnarray}
Here ${\bf a}$ and ${\bf b}$, the components of ${\bf k'}$ and ${\bf (k-k')}$
perpendicular to the line of sight, are two dimensional
vectors in the plane of the sky and hence their cross product is a
(pseudo)scalar quantity.  Because we obtained these expressions from a
convolution, they should be symmetric under the interchange of
${\bf k'}$ and ${\bf (k-k')}$, and this can be easily verified.  It is
also clear that the $E$-mode is a scalar and the $B$-mode is a pseudo-scalar.

The power spectrum is defined by
\begin{equation}
  \langle \epsilon({\bf k})\epsilon^{*}({\bf k'}) \rangle =
   (2 \pi)^{3} \delta({\mathbf k-k'})P_{\epsilon\epsilon}(k) ,
\end{equation}
and is given by
\begin{eqnarray}
 \langle \epsilon({\mathbf k_1}) \epsilon^{*}({\mathbf k_2}) \rangle
  k_1^2 k_2^2
 &=& {1\over 225} (2\pi)^{3} \delta({\mathbf k_1-k_2}) 
\int {d^3{\mathbf k'}\over (2\pi)^3} P_{\Phi}(k')P_{\Phi}(|{\mathbf k_{1}-k'}|)
      \nonumber \\
 &\times& f_{\epsilon}(\kpp,\kp -\kpp,k_{z}') 
  \left[ f_{\epsilon}(\kpp,\kp -\kpp,k_{z}') +
  f_{\epsilon}(\kp -\kpp,\kpp,-k_{z}') \right] \,,
\end{eqnarray}
and similarly for the $B$-mode.
Because both expressions are symmetric under the interchange of $\bf{k'}$
and ${\mathbf(k-k')}$ though, the power spectrum becomes somewhat simpler:
\begin{equation}
 \langle \epsilon({\mathbf k_{1}}) \epsilon^{*}({\mathbf k_{2}}) \rangle
k_{1}^{2} k_{2}^{2}
 = \frac{1}{225} (2\pi)^{3} \delta({\mathbf k_{1}-k_{2}}) 
   \int \frac{d^{3}{\mathbf k'}}{(2 \pi)^{3}} 
   P_{\Phi}(k')P_{\Phi}({\mathbf |k_{1}-k'|})
   2 f_{\epsilon}^{2}\left( \kpp,(\kp -\kpp),k'_{z}\right) \,.
\end{equation}

We now need to evaluate the functions $f_{\epsilon,\beta}$ and perform
the convolution integral.  Since we only consider ${\bf k}$
perpendicular to the line of sight, we choose our Cartesian axes such
that ${\bf k} = k(1,0,0)$, and we choose polar angles defined so that
${\bf k'} = \alpha k
(\cos{\theta},\sin{\theta}\cos{\phi},\sin{\theta}\sin{\phi})$ with
$\alpha \equiv k/k'$.  In this coordinate system, $|{\mathbf k-k'}| = k
\sqrt{1+\alpha^{2}-2\alpha\mu}$ (where $\mu = \cos{\theta}$) with no
dependence on $\phi$.  We find that
\begin{eqnarray}
 \left[ f_{\epsilon}(k,\alpha, \mu, \phi)\right]^{2} &=&  k^{12}\alpha^{4} 
       \left[ (\alpha^{2} - (1+2\alpha^{2})\mu^{2} + 2\alpha\mu^{3}
       + (1-\alpha^{2}+\alpha\mu)(1-\mu^{2})\cos{\phi}^{2}) \right]^{2}
\nonumber \\
 \left[ f_{\beta}(k,\alpha, \mu, \phi)\right]^{2} &=& 
k^{12} \alpha^{4} (1-\mu^{2})(\alpha-\mu)^{2}(1-2\alpha\mu)^{2}\sin{\phi}^{2} 
\,.
\end{eqnarray}
Integrating over $d\phi/(2\pi)$ gives functions which are polynomials in
$\mu$ and $\alpha$.  Taking a factor of $k^{12}\alpha^{4}$ out of the
polynomial and denoting the remainder by $g_{\epsilon\epsilon}(\alpha,\mu)$
and $g_{\beta\beta}(\alpha, \mu)$ we find
\begin{eqnarray}
 g_{\epsilon\epsilon}(\alpha, \mu) &=&
  {1 \over 8}( (1+ \alpha^4)(3-14\mu^2 +19\mu^4)+
  2\alpha(3\mu -2\mu^3 -17\mu^5) 
\nonumber \\
& &+ \alpha^2(2 -17\mu^2 +44\mu^4 +19\mu^6)+2\alpha^3(\mu +2\mu^3 -19\mu^5)) \,,
\nonumber \\
 g_{\beta\beta}(\alpha, \mu) &=& {1 \over 2}(\alpha -\mu)^2 (1-2\alpha\mu)^2
  (1-\mu^{2}) \,.
\end{eqnarray}
Now the power spectrum is given by
\begin{equation}
P_{\epsilon\epsilon}(k) = {2k^8\over 225}
  \int {k^3 \alpha^2 d\alpha d\mu \over (2\pi)^2}
   P_{\Phi}(\alpha k) P_{\Phi}(k \sqrt{1+\alpha^2 - 2\alpha \mu}\,) \alpha^4 
   g_{\epsilon\epsilon}(\alpha,\mu)
   + k_{||}{\rm -terms} \,.
\end{equation}
where $k_{||}$-terms indicates terms with line-of-sight component $k_{||}\ne 0$
which will not contribute to our final result.
The cross power spectrum $P_{\epsilon\beta}(k)$ is identically zero as it 
is parity violating.
We use Poisson's equation, and express the ellipticity power spectrum in 
dimensionless form as
\begin{equation}
  \Delta^{2}_{\epsilon\epsilon}(k) =
  {1 \over 225} \left({3 \over 2} \Omega_{0} H_{0}^2\right)^4
  \int {d\alpha \over \alpha}\ \Delta^{2}_m(\alpha k)
  \int_{-1}^{1} d\mu\ {\Delta^{2}_m(k\sqrt{1+\alpha^2-2\alpha\mu})\over
  (1+\alpha^2 -2\alpha\mu)^{7/2}}\ g_{\epsilon\epsilon}(\alpha,\mu)
  + \cdots \,,
\label{eqn:pkee}
\end{equation} 
and similarly for the $B$-mode power spectrum.
Here $\Delta^2_m$ is the contribution to the mass variance per $\log k$
\begin{equation}
  \Delta^2_m(k) \equiv {d\sigma^2\over d\log k} = {k^3 P(k)\over 2\pi^2} \,.
\end{equation}

To proceed to the angular power spectrum we use the Limber approximation.
In Fourier space the Limber approximation states that the angular power
spectrum of a projection, $\int d\chi\ w(\chi)S$, of a scalar field
$S({\bf x})$, is \cite{kai92,WhiHu00}
\begin{equation}
  {\ell (2\ell+1)C_\ell\over 4\pi} = {\pi\over\ell} \int \chi d\chi
  \ w^2(\chi) \Delta^2_{SS}(k\chi=\ell,a) \,,
\label{eqn:limber}
\end{equation}         
where $w(\chi)$ is the weight function at radial distance $\chi$.
Before we compute our final result we need to return to the issue of the
normalization constant $C$.

\subsection{Normalisation} \label{sec:normalization}

We determine $C$ empirically by fitting the model prediction for the
mean-square source ellipticity to that  which is observed,
\begin{eqnarray}
  \bar{\epsilon}^2 &\equiv&  \langle \gamp^{2} + \gamx^{2} \rangle 
\nonumber \\
 &=& C^{2} \langle (L_{x}^{2} + L_{y}^{2})^{2} \rangle \, .
\end{eqnarray}
We are assuming here that all of a galaxy's ellipticity is due to its 
angular momentum acquired via tidal torques.  It is likely that some 
fraction of the ellipticity is due to other effects (e.g.~halo shapes as in 
CKB) in which case our  normalisation provides an upper limit
to the angular momentum correlations.

Averaging over all orientations of ${\bf L}$ we find
\begin{equation}
 \bar{\epsilon}^2 = {8 C^{2} \over 15} \langle L^{4} \rangle
 = {24 C^{2} \over 15} \langle L^{2} \rangle^{2} \,.
\end{equation}
The expectation value of $L^2$ is calculated \cite{CatThe96} to obtain
\begin{equation}
  \bar{\epsilon}^2 = {32 C^2\over 1125}\left( {3\over 2}\Omega_mH_0^2\right)^4
  \sigma^4(R),
\label{eqn:norm}
\end{equation}
where all the time dependent terms and the magnitude of the inertia
tensor have been absorbed into the constant $C$.  Note $C \propto
1/\langle L^2 \rangle$ so any time dependence in $L$ cancels out in the expressions
for $\gamma_{i}$. We can therefore calculate everything according to
linear theory extrapolated to the present ($a=1$).  
Also, the magnitude of $L$ is irrelevant in our calculation;
what is important is the ratio of $L^2$ to the mean squared value 
$\langle L^2 \rangle$.
The quantity,
\begin{equation}
  \sigma^2(R) \equiv \int {dk\over k}
     \ \Delta^2_m(k,a=1) \left({3j_1(kR)\over kR}\right)^2 \,,
\end{equation}
is the variance of the density field smoothed on a scale $R$,
associated with the size of a galaxy mass region in the initial
density field.  We will take $R=1h^{-1}$Mpc , but calculate
the dependence of our results on this choice in \S\ref{sec:amplitude}.
Finally we obtain the normalised ellipticities, 
\begin{eqnarray}
\gamp &=& \sqrt{{1125\over32}} {\bar{\epsilon} \, (L_{x}^{2}-L_{y}^{2}) \over
             \left( {3\over 2}\Omega_mH_0^2\right)^{2} \sigma^{2}(R)} \,,
\nonumber \\
\gamx &=& \sqrt{{1125\over32}} {\bar{\epsilon} \, 2 L_{x}L_{y} \over
             \left( {3\over 2}\Omega_mH_0^2\right)^{2} \sigma^{2}(R)} \,.
\end{eqnarray}
We take $\bar{\epsilon} = 0.4$ to present our results, as it is a good fit
to the observed rms galaxy ellipticity (see CNPT).

\subsection{Summary} \label{sec:resultsummary}

We are now in a position to derive our central result.
Including the normalisation factors from \S\ref{sec:normalization}, we
obtain the angular power spectrum 
\begin{equation}
 {\ell(2\ell+1)C^{\epsilon\epsilon}_\ell\over 4\pi} = 
 {5\bar{\epsilon}^2 \over 32\sigma^4(R)}{\pi \over \ell} 
 \int \chi d\chi \ w^2(\chi) 
 \int {d\alpha \over \alpha} \Delta^{2}_m({\alpha \ell \over \chi})
 \int_{-1}^{1} d\mu {\Delta^{2}_m(\ell
 \sqrt{1+\alpha^2-2\alpha\mu}/\chi) \over (1+\alpha^2
 -2\alpha\mu)^{7/2}} g_{\epsilon\epsilon}(\alpha,\mu) \;,
\label{eq:l2cl}
\end{equation}
and similarly for the $C^{\beta\beta}_\ell$ power spectrum.
The cross-spectrum vanishes due to parity.

In integrating equation~(\ref{eq:l2cl}) the factor $(1+\alpha^2 -2\alpha\mu)^{-7/2}$
must be treated carefully near $k \gg 1$ and $k \simeq k'$.
We make an approximation in this region which is good to $\le 7$ per cent in
the worst case (a low redshift galaxy distribution and $\ell > 10^3$).
The weight function depends on the number of sources per unit redshift.
We take
\begin{equation}
  {dn\over d\chi} \propto \chi^\alpha \exp\left[
  -\left( {\chi\over \chi_*} \right)^\beta \right]\,,
\end{equation}
with $\int dn=1$.  The case $\alpha=1$, $\beta=4$ approximates the
source distribution of a flux limited survey, and we adjust $\chi_*$
to obtain a desired mean source redshift (see below).

It is straightforward to go from the angular power spectrum to the
correlation function.  The two are related by a slight generalization
of the Hankel transform as
\begin{eqnarray}
\left\langle \gamma_1 \gamma_1 \right\rangle &=& \int {\ell d\ell \over 4\pi}
  ( C_{\ell}^{\epsilon\epsilon}
  \left[ J_0(\ell \theta) +J_4(\ell \theta)\cos{4\phi}\right]
 + C_{\ell}^{\beta\beta}
  \left[ J_0(\ell \theta) -J_4(\ell \theta) \cos{4\phi}\right] )\,,  
\nonumber \\
\left\langle \gamma_2 \gamma_2 \right\rangle &=& \int {\ell d\ell\over 4\pi}
  \left( C_\ell^{\epsilon\epsilon}
         \left[J_0(\ell \theta) -J_4(\ell \theta)\cos{4\phi}\right]
   + C_\ell^{\beta\beta}
         \left[ J_0(\ell \theta) +J_4(\ell \theta)\cos{4\phi}\right] \right)\,,
\label{eqn:corr}
\end{eqnarray}
where the separation between the two points has polar coordinates
$(\theta,\phi)$.
For comparison with observations we also calculate the ``ellipticity 
variance'' smoothed on some angular scale $\theta$. This is obtained by 
convolving the ellipticities $\gamma_{i}$ with a filter function and 
taking the autocorrelation of this at zero lag. For a top-hat real 
space circular filter of radius $\theta$, the variance is given by
\begin{equation}
  \sigma^2(\theta) = \int {\ell d\ell \over 2 \pi} 
    ( C_\ell^{\epsilon\epsilon} + C_\ell^{\beta\beta} )
    \left[ {2 J_1(\ell\theta) \over \ell\theta} \right]^2 \;.
\label{eqn:var}
\end{equation}

\section{Results} \label{sec:results}

\begin{figure*}
\begin{center}
\leavevmode
\epsfxsize=7in \epsfbox{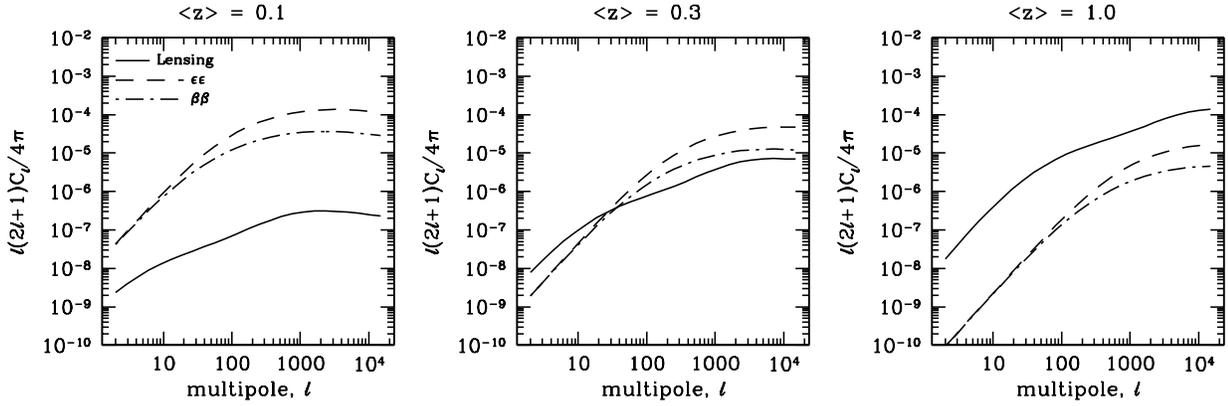}
\end{center}
\caption{The various angular power spectra discussed in the text for
our flat $\Lambda$CDM model.  The solid line is the predicted weak
lensing signal , the dashed line the ``intrinsic'' ellipticity $EE$
power spectrum and the dot-dashed line the intrinsic $BB$ power
spectrum.  The first panel is for a low redshift source galaxy
distribution of mean redshift $\langle z \rangle = 0.1$ using the
redshift distribution described in the text.  The second and third
panels are for $\langle z \rangle =0.3$ and 1.0 respectively.}
\label{fig:l2cl}
\end{figure*}

To demonstrate our results we use a flat $\Lambda$CDM cosmological model with 
parameters $(\Omega_m,\Omega_b h^2, \Omega_{\Lambda}, h, n, \sigma_{8}) = 
(0.3, 0.018, 0.7, 0.67, 1.0, 0.9) $, 
with the fit to the transfer function of Eisenstein \& Hu~\shortcite{eishu99}.
Fig.~\ref{fig:l2cl} shows the ellipticity-ellipticity power spectra, along
with the power spectrum predicted by weak lensing for three source
distributions (with mean source redshifts $\langle z_{\rm src}\rangle=0.1$,
0.3, and 1.0).

\subsection{Amplitude of the angular power spectra} \label{sec:amplitude}

We first discuss the overall amplitude of the intrinsic and lensing
power spectra.  The scales of most relevance for recent measurements
of intrinsic correlations \cite{brownetal00} and for current weak
lensing surveys are from $\ell \sim 100$ ($\theta \sim 2^{o}$) to
$\ell\sim 1000$ ($\theta \sim 10'$).  For our low redshift galaxy
sample with $\langle z_{\rm src}\rangle=0.1$ the intrinsic signal
dominates the weak lensing signal, while the situation is reversed for
the higher redshift samples.  This is one of our main results, and
agrees with calculations and measurements by previous authors (CKB,
CNPT, Heavens et al. 2000, Croft \&
Metzler 2000, Brown et al. 2000).
Our result adds further support that this is the order of magnitude to
expect intrinsic correlations due to correlations in the tidal field.

The result can be easily understood as the combination of two effects
(these were also alluded to by Miralda-Escud\'e~\shortcite{mir91} as
an argument for ignoring intrinsic alignments in deep surveys).  For
lensing the power spectrum is proportional to the square of the
projected mass density, which increases with increasing survey depth.
Intrinsic correlations, however, are located in the source plane due
to galaxies physically close to each other.  For deeper surveys, the
probability of two galaxies close to each other on the sky being
physically close to each other is much smaller than for shallow
surveys.  So the intrinsic correlations get `washed out' by this
projection effect.

This result shows that for deep weak lensing surveys with $\langle
z_{\rm src}\rangle \sim 1.0$, intrinsic correlations are unlikely to
contribute to the signal at a significant level, as long as the 
redshift distribution of sources is broad.  For shallower surveys, 
however, they are expected to be the dominant signal.
This has been convincingly demonstrated by Brown et
al.~\shortcite{brownetal00} using the SuperCOSMOS Sky Survey.  They
detect a significant intrinsic correlation of the order of magnitude
of our results.  We compare our results with these observations in
more detail in the next section.

There are two main uncertainties in the amplitude of our prediction.
The first is that for very large $\ell$, we expect non-linear clustering of
galaxies to erase the initial correlations due to interactions with
neighbouring galaxies.  This may randomise ellipticities altogether, or
it may create new correlations with a different amplitude and scale 
dependence.  Also, based on numerical results, the overall amplitude
of the signal may be scaled down by some factor, due to the weakening
of spin-spin correlations over time.  
Secondly, the smoothing scale in our normalisations affects our
results.  We use $R=1 h^{-1}$Mpc as the comoving scale containing on
average a galaxy mass ($10^{11}M_{\odot}$) at the background density.
Our results scale as $\sigma^{-4}(R)$, so changing the smoothing scale
alters things significantly.  For our cosmological model, with
$\sigma_8=0.90$, we find that $\sigma(R)=3.6$, 2.8 and 2.0 for
$R=0.5$, 1.0 and $2.0 h^{-1}$Mpc respectively.  With these numbers,
increasing (decreasing) our smoothing scale by a factor of two
increases (decreases) our results by a factor of $3.5$ ($2.9$).

The rms ellipticity of galaxies empirically fixes the normalisation.  
We choose $\bar{\epsilon}= 0.4$ as it is unlikely to be larger than this, 
but it may be somewhat smaller.
As stated earlier, it is also possible that only a fraction, $f < 1$, of 
the rms ellipticity is 
determined by the angular momentum, with the rest contributed by other 
processes (e.g. halo shapes).  In this case our results should be multiplied
by this factor of $f$, thus lowering the predicted power spectrum.  We 
show all our results assuming $f=1$ to emphasize how large intrinsic 
correlations could be.

\subsection{Power Spectrum Shape}

The shape of the intrinsic power spectra is similar for all three redshift
distributions, and is reflective of the underlying physics.
On large scales (small $\ell$) the (log) slope is 2 for both the $EE$ and $BB$
power spectra, consistent with shot noise
($C_{\ell}^{\epsilon\epsilon}=C_{\ell}^{\beta\beta}=$constant).
This is expected as there should be no intrinsic correlation between sources
that are widely separated on the sky.
The lensing signal falls off more slowly at large angles because it is an 
integral over the power spectrum whereas the intrinsic signal is an 
integral over the square of the power spectrum.  Also lensing occurs 
between us and the source galaxies, whereas intrinsic correlations
are at the source.

The power begins to fall significantly below the shot noise power on an
angular scale corresponding roughly to the turnover in the mass power spectrum
at the mean source redshift.  
The linear theory power spectrum changes slope at $k\sim 0.02 h\, {\rm Mpc}^{-1}$, 
and $k\chi \sim \ell$ at the mean redshift, so we can work out what value
of $\ell$ we expect this feature in $\Delta^2_m(k)$ to show up.
Mean redshifts of $z=0.1$, $0.3$ and $1.0$ correspond to distances of
$\chi = 280$, 730 and $1650h^{-1}$Mpc in our cosmological model.
Thus we expect the roll over to occur at roughly $\ell\sim 6$, 15 
and 35 for the three source distributions in Fig.~\ref{fig:l2cl}.
This is indeed what we see in that the $E$- and $B$-modes start to differ 
from each other and to change slope for larger values of $\ell$.

The large $\ell$ slope flattens off (and may even start to decrease for the
low redshift population) due to the fact that the 3D matter power spectrum
becomes flat at large $k$.
We emphasize again, however, that on very small angular scales our predictions
may be completely overwritten by non-linear clustering effects.

\subsection{Ratio of $E$- to $B$-mode power} \label{sec:ebmodes} 

The $E$-mode intrinsic signal is enhanced over the $B$-mode by a factor of
$C_{\ell}^{\epsilon\epsilon} / C_{\ell}^{\beta\beta} \simeq 3.5$ on small 
scales ($\ell \gg 1$).  This goes to a constant ratio for all 
three galaxy redshift distributions, just shifted to smaller angular scales 
for deeper surveys.
On very small scales the power spectrum becomes roughly constant (increases 
only as $\log{k}$), so this ratio must be determined by the geometry of the
problem and the tensor nature of the source.  A similar result was obtained 
by \cite{HuWhi97} in studying the polarisation induced in the CMB by tensor 
perturbations (gravitational waves), where they found that the ratio tended
to $C_{\ell}^{\epsilon\epsilon} / C_{\ell}^{\beta\beta} = 13/8$ in the 
small angle limit.
Given the mathematical similarity between polarization and galaxy
ellipticities, it is likely that similar geometrical considerations
are producing the $B$-mode suppression in our model, although the suppression
is stronger in our case.

An intuitive way of thinking about this $B$-mode suppression is that
an isolated point mass can generate only $E$-modes.  We believe that this
is the reason our $B$-modes are suppressed on smaller scales where the
density field can be increasingly described in terms of ``objects''.
As shown by Lee \& Pen~\shortcite{LeePen00b}, if the
tidal field has ordered eigenvalues $\lambda_{1} \geq  \lambda_{2} \geq
\lambda_{3}$ along its principal axes then (in this principal axes
frame) $L_{1} \propto
(\lambda_{2}-\lambda_{3})I_{23}$ and similarly for the other components.
If ${\bf I}$ and ${\bf T}$ are uncorrelated we expect $L_2$ to by largest from the 
ratio of the sizes of the eigenvalues.  Porciani, Dekel \&
Hoffman~\shortcite{PorDekHof01b} showed that in their simulations, in general
$L_1$ is smaller than the other components, so the angular momentum is
largely in the plane perpendicular to the first principal axis of ${\bf T}$.
This can be understood by
considering a proto-galaxy forming in the vicinity of a large point mass (or 
equivalently a large spherically symmetric mass distribution).  In the limit that
the tidal field is dominated by this mass, the first principal axis of ${\bf T}$ will
lie in the radial direction (i.e. along the separation vector between the 
proto-galaxy and the point mass), and the second and third will be in the 
tangential directions.  The second and third eigenvalues are identical,
 because there is no preferred tangential direction, making $L_1=0$.
Thus the angular momentum acquired due to tidal torques will be 
perpendicular to its separation vector from the point mass, and 
hence the galaxy will be oriented radially.  This is a pure 
$E$-mode pattern.  To see this in
the context of our calculation, notice that only the radial first
derivative of $\Phi$ is non-zero, so all cross derivatives
$\Phi_{,ij}$ (with $i \neq j$) vanish.  These cross derivatives become
cross products of $k$-modes in Fourier space, so inspection of
equation~(\ref{eqn:ebmodefunctions}) shows that the $B$-mode must
vanish.  So if the shear field is dominated by isolated point masses,
then the correlations will be mostly $E$-modes.

That our model produces $B$-modes at all is due to the ellipticity
components $\gamma_{i}$ being quadratic in ${\bf L}$ and hence in
$\Phi({\mathbf x})$.  This can be seen in our expression for
$\beta({\mathbf k})$ in equations~(\ref{eqn:ebmodes}) and
(\ref{eqn:ebmodefunctions}) as a convolution over the Fourier modes of
the potential.
Examination of equations (\ref{eqn:ebmodes},\ref{eqn:ebmodefunctions})
shows that most of the contribution comes when ${\mathbf k}$ and
${\mathbf k'}$ are quite different in magnitude.
We interpret this as indicating that large-scale perturbations on the
small-scale potential field generate the $B$-modes.

One of the main ways of discriminating between weak lensing and
systematic effects is that weak lensing produces only $E$-mode power.
Shot noise has $C_{\ell}^{\epsilon\epsilon} = C_{\ell}^{\beta\beta} =
{\rm constant}$, so looking at the $B$-modes gives an estimate of the level
of noise in the $E$-mode lensing signal.  Our model indicates that
there can be significantly larger intrinsic $E$-mode power than would
be naively be expected from looking at the level of $B$-mode
power.  Thus the suppression of the $B$-modes in our model means that
intrinsic correlations may be hidden in a low signal-to-noise
detection of $E$-mode power.  On the other hand, this factor of $3.5$
suppression should be an upper limit to a real measurement.  Adding
shot noise, with $C_{\ell}^{\epsilon\epsilon} =
C_{\ell}^{\beta\beta}$, will only reduce the $E$- to $B$- ratio.
Non-linear effects adding small-scale correlations may also alter the
ratio.

\begin{figure*}
\begin{center}
\leavevmode
\epsfxsize=7in \epsfbox{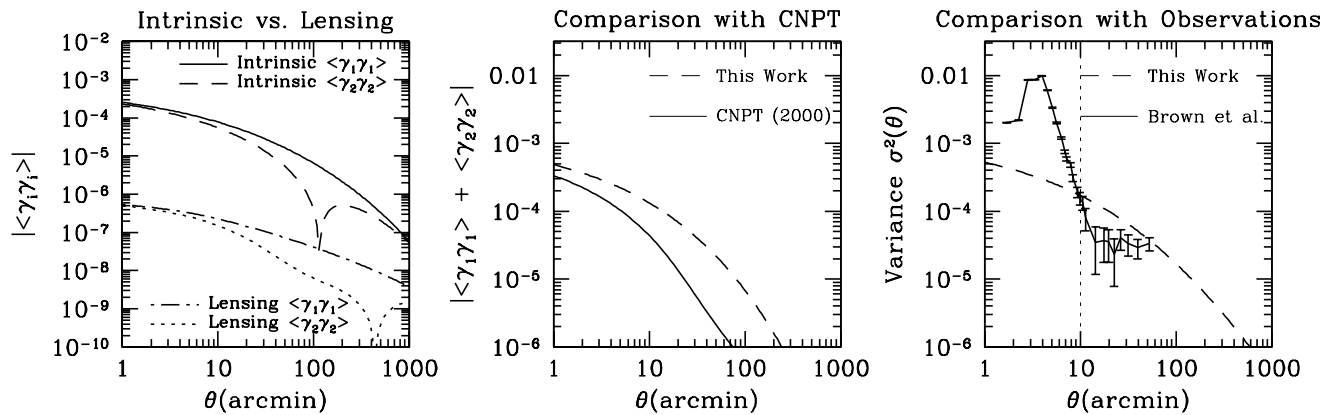}
\end{center}
\caption{
Correlation functions and comparison with other work, all for the low 
redshift $\langle z \rangle =0.1$ source distribution.  Note that different
quantities are plotted on each panel, with a different y-axis scale on the 
first one.
Left panel: Comparison of the correlation functions for intrinsic 
ellipticity correlations (in our model) and lensing induced correlations, 
as calculated from the power spectra in Fig. 1.  
Middle panel: Comparison with the theoretical results of CNPT. Plotted
is the sum of the two correlation functions in each case.  Curve is 
from Fig.~5 of CNPT for $a=0.24$.  Note the similarity of our results, 
although the amplitude of our result is slightly larger.
Right panel: Comparison with the observations of Brown et al. (2000).  
Plotted is the ellipticity variance as described in the text.  The
vertical line at $10'$ corresponds to a linear scale of roughly $1h^{-1}$Mpc
at the mean source redshift, below which we do not expect our calculation
to be relevant.}
\label{fig:compare}
\end{figure*}

\section{Discussion and Comparison with other work} \label{sec:comparison}

\subsection{Correlation Functions and Comparison with CNPT and Simulations}

We calculated correlation functions from the power spectra in
Fig.~\ref{fig:l2cl} using equation~(\ref{eqn:corr}) for the low redshift source
distribution.  These are plotted in the left panel of Fig.~\ref{fig:compare}.  
In evaluating equation~(\ref{eqn:corr}) we use $\phi=0$, corresponding to the 
separation vector of the two galaxies being along the $x$-axis.
Note that the $\langle \gamx\gamx \rangle$
function goes anticorrelated for large separations, and so we plot its
absolute value.  The physics producing the curves is less clear than
in the power spectrum analysis, but they do show the same basic features ---
intrinsic correlations should dominate weak lensing signal at low redshift.

Computing correlation functions allows us to compare our results with
the work of CNPT who pursued a real space correlation function analysis 
of correlations in the orientations of galaxies.  This
comparison is shown in the middle panel of Fig.~\ref{fig:compare},
where we are comparing with the $a=0.24$ curve in Fig. 5 of CNPT.  
\footnote{Note that CNPT seem to define `$a$' differently to Lee \& 
Pen~\shortcite{LeePen00a}, so that the $a=0.24$ they plot (and which 
we compare to) may not correspond exactly to the $a=0.24$ found by Lee \&
Pen in their simulations.}
The
first thing to note is that our results are broadly similar to theirs.
The amplitude of our signal is slightly higher, although increasing
the value of the correlation parameter `$a$' in their model can
reverse this. 
This shows that extrapolating their model to $a=1$ (for
the case of uncorrelated ${\bf I}$ and ${\bf T}$ tensors) overestimates the
correlations.  The reason for this is an approximation made in their
calculation which assumes $a \ll 1$, so it was not designed to 
work when $a=1$.  It is interesting that our results only
differ by $\sim 50$ per cent at $\theta \sim 1'$.
We would expect, because they normalise their model to take account
of the suppression of spin-spin correlations seen between dark halos
in numerical simulations, that their result should be significantly 
smaller than ours.  It may be that differences in the normalisation of
the underlying dark matter power spectrum, or other model differences,
are decreasing the difference between our results.  Overall, given the
major simplifications and assumptions that go into each model, we feel
it is encouraging that both models obtain similar results.  

The slopes of our correlation functions are also somewhat different,
with our correlation function falling off more slowly at large angular
separations.  We suspect that this is probably reflecting the
different matter power spectra used in the calculations (CNPT used a
power law whereas we use the full linear-theory power spectrum
appropriate for our cosmological model), but it may also be due to
differences in our models.  The fact that our results are so similar
adds support to the calculation of CNPT, and suggests that both their
results and ours are reflecting the physics of tidal torques and not
the details of our respective models.

This conclusion is further supported by recent numerical work. 
Heavens et al.~\shortcite{HeaRefHey00} assumed a galaxy disk will form 
perpendicular to the angular momentum vector of the halo it lives in
 and calculated spin--spin correlations from
large numerical simulations.  They found significant spin correlations
for halos at $z=1$, and less significant correlations at low redshift.
At the same time, Croft \& Metzler~\shortcite{CroMet00} measured halo
shape correlations in similar large simulations and found correlations
of similar magnitude to (but slightly stronger than) the spin correlations of
Heavens et al.~\shortcite{HeaRefHey00}.  
Our results are in rough agreement with the amplitude of the signal found
in these numerical simulations.

\subsection{$E$- and $B$-Modes}

In a second paper, Crittenden et al.~\shortcite{cnpt00b} calculate $E$- 
and $B$-mode 
correlation functions by constructing local estimators of the 
$E$- and $B$-mode contributions to the ellipticities in real
space.  
While their method of analysis is very different to the power spectrum
approach we have presented here (see Hu \& White~\shortcite{HuWhi00}
for an application of this method to simulated weak lensing data),
the calculations are mathematically equivalent.  
We find that our results are qualitatively different from each other.
For two different matter correlation functions (power laws with slopes
$-1$ and $-3/2$) they find that the $E$- and $B$-mode correlation functions 
are the same for small separations in both cases.  This is to be contrasted
with the power spectra we obtain in Fig.~\ref{fig:l2cl}, where the two are
the same on large scales and differ on small scales.  They also find that
on large scales, for $\xi(r) \propto r^{-1}$, the $E$-modes are enhanced
somewhat over $B$-modes, while they are identical for 
$\xi(r) \propto r^{-3/2}$.  The fact that the relative contributions 
change depending on the power law slope of the correlation function 
indicates that this is the source of the difference between their results
and ours.  We use a power spectrum which has different large and small 
scale slopes to the corresponding slopes in their correlation functions,
so our resulting $E$- and $B$-modes look different in the large and small
angle limits.
We have investigated the power spectrum dependence of the $E/B$ ratio by 
considering `tilted' power spectra with scalar perturbation spectral indices of 
$n = (0.75,1.25)$ as well as the canonical $n=1$ which we used to 
present our results.  We find that indeed the ratio (at $\ell =5000$ for 
$\langle z\rangle =0.1$) decreases with increasing spectral index.  A $25\%$
change in $n$ produces roughly a $12\%$ change in the $E/B$ ratio.
The reason for this is alluded to in \S\ref{sec:ebmodes}, where we showed
that for a given value of $\ell$, the contributions to the $E$- and $B$-modes 
are from a range of different scales, and the ranges are different for the 
two modes.  Tilting the power spectrum changes the ratios of power on 
different scales, and so will change the relative contributions to the $E$- 
and $B$-modes.

\subsection{Comparison with Observations}

In the right panel of Fig.~\ref{fig:compare} we plot the intrinsic 
ellipticity variance as calculated from equation~(\ref{eqn:var}) together with
the observations of Brown et al.~\shortcite{brownetal00}.
Observations by Pen et al.~\shortcite{penleesel00} are at a similar level, but
at lower statistical significance and over a smaller range of scales, so we
only compare to Brown et al.\ here.
At first sight it appears that our calculation does not fit the data very 
well.  It is encouraging, however, that we have the right order of magnitude, 
and there are a number of other factors to take into account.
At redshift $z=0.1$, $10$ arcminutes (indicated by the vertical line in the
figure) projects onto a linear separation of 
$\sim 1 h^{-1}$Mpc.  We expect that non-linear clustering and dynamical 
interactions will have erased initial conditions for galaxy separations 
much smaller than this.  Non-linear effects may produce their own alignments, 
and we suspect this is what is giving the large observed signal at small 
separations.

On larger separation scales, we have argued that our prediction should
be an upper limit to correlations due to tidal torques.  So the fact
that we have obtained the right order of magnitude and are on the high
side of the data is a good sign.  The agreement of our results and the
data on angular scales from $10'$ to $100'$ is really quite good,
taking this into account.  In order to conclusively test whether this
observed correlation is due to correlations in the tidal field we need
data that goes out to larger scales.  It is important to see whether
the variance stays constant as the data hint at, or whether it falls
off on larger scales as the theory predicts.

\section{Conclusions} \label{sec:conclusions}

In conclusion, we have presented a new analytic calculation of ellipticity 
correlations due to correlations in the tidal field galaxies form in.
This correlation is probably giving the observed signal seen by 
\cite{brownetal00} on scales of $10'$ to $100'$.
Our calculation is complementary to previous work of CNPT, who performed 
a similar calculation with similar input physics but different assumptions
relating the physics to the observed ellipticity.
The similarity of our results adds support to their validity, and indicates
that they may be reflecting the physics of tidal torques rather than
the detailed model assumptions.

We extend previous work in applying the Fourier space $E$- and
$B$-mode power spectrum formalism to the problem (but see Crittenden
et al.~\shortcite{cnpt00b} for an application of this decomposition in
real space).  We find that $E$-modes are enhanced by a factor of
$\simeq 3.5$ over $B$-modes on small scales in our model, which can be
understood intuitively by noting that isolated point masses can
generate only $E$-modes.  This means intrinsic $E$-mode contamination
of weak lensing signal can be considerably larger than that implied by
looking at the $B$-modes.  This result differs from the findings of
Crittenden et al.~\shortcite{cnpt00b}, who found that $E$- and $B$-modes
have the same amplitude on small scales.  This is probably mostly reflecting
the different matter power spectra used in the calculations, but may also
be at least partly due to differing model assumptions.

We confirm the results of previous authors in finding that intrinsic
correlations are a small but possibly significant contaminant for weak
lensing surveys with mean redshifts of order unity.  We have shown that
the contamination can be as large as $30 \%$ (in rms correlation) on
degree to 10-arcmin angular scales, but this should be an upper limit
according to our calculation.
Thus, while at present uncertainties due to intrinsic shape correlations
are at most comparable to (and probably smaller than) other observational
uncertainties for deep surveys, they will become important for future
precsion surveys designed to put strong constraints on cosmological
parameters.
They will also be important for surveys in which the galaxy sample is
distributed into narrow redshift bins by photometric techniques.  Croft
\& Metzler~\shortcite{CroMet00} have shown that reducing the width of the
redshift distribution increases the intrinsic signal while leaving the
lensing signal essentially unchanged.

Shallow surveys, on the other hand, are a very good probe of intrinsic
correlations.  These correlations should be investigated in more detail 
because they are interesting in their own right.  
Their existence and strength may be able put constraints on models of
galaxy formation, and on the relationship between baryonic matter and 
dark matter halos by further comparison with numerical simulations and 
analytic theory.

\section*{Acknowledgments}
We thank U-L. Pen, C. Metzler, R. Croft, R. Crittenden, P. Natarajan, 
A. Heavens and A. Refregier for useful discussions on intrinsic
alignments.
JM thanks R. Narayan and J. Huchra for helpful comments on an earlier
draft.
This work was supported in part by the Alfred P. Sloan Foundation and the
National Science Foundation, through grants PHY-0096151, ACI96-19019 and
AST-9803137.
MK was supported at Caltech by NSF AST-0096023, NASA
NAG5-8506, and DoE DE-FG03-92-ER40701.
We are grateful to Lindsay King for pointing out an error in our lensing
calculation.

{}


\begin{thebibliography}{}

\bibitem[\protect\citename{Abel, Croft \& Hernquist, }2001]
    {AbeCroHer01}
    Abel T., Croft R., Hernquist L., 2001, [astro-ph/0111046]

\bibitem[\protect\citename{Bacon, Refregier \& Ellis }2000]{BacRefEll00}
    Bacon D., Refregier A., Ellis R. S., 2000, MNRAS, 318, 625

\bibitem[\protect\citename{Bartelmann \& Schneider }1992]{bar92}
    Bartelmann M. \& Schneider P., 1992, A\&A, 259, 413

\bibitem[\protect\citename{Bartelmann \& Schneider }1999]{bar99}
    Bartelmann M. \& Schneider P., 1999, Physics Reports, 340, 291 

\bibitem[\protect\citename{Blandford et al. }1991]{bla91}
    Blandford R. D., Saust A. B., Brainerd T. G., Villumsen J. V., 1991, MNRAS,
     251, 600

\bibitem[\protect\citename{Brown et al. }2002]{brownetal00}
    Brown M. L., Taylor A. N., Hambly N. C., Dye S., 
     2002, MNRAS, in press, [astro-ph/0009499]

\bibitem[\protect\citename{Cabanela \& Aldering }1998]{CabAld98}
    Cabanela J. E. \& Aldering G., 1998, AJ, 116, 1094

\bibitem[\protect\citename{Catelan, Kamionkowski \& Blandford }2000]{CatKamBla00}
    Catelan P., Kamionkowski M., Blandford R.D., 2000, MNRAS, 320, 7 (CKB)

\bibitem[\protect\citename{Catelan \& Porciani }2001]{CatPor01}
    Catelan P. \& Porciani C., 2001, MNRAS, 323, 713

\bibitem[\protect\citename{Catelan \& Theuns }1996]{CatThe96}
    Catelan P. \& Theuns T., 1996, MNRAS, 282, 436

\bibitem[\protect\citename{Crittenden et al. }2001]{cnpt00a}
    Crittenden R. G., Natarajan P., Pen U., Theuns T., 2001,
    ApJ, 559, 52 (CNPT)

\bibitem[\protect\citename{Crittenden et al. }2000]{cnpt00b}
    Crittenden R. G., Natarajan P., Pen U., Theuns T., 2000,
    [astro-ph/0012336]

\bibitem[\protect\citename{Croft \& Metzler }2000]{CroMet00}
     Croft R. \& Metzler C., 2000, ApJ, 545, 561


\bibitem[\protect\citename{Djorgovski }1987]{Djo87}
     Djorgovski S., 1987, in {\em Nearly Normal Galaxies}, (Springer, New York), p.227

\bibitem[\protect\citename{Doroshkevich }1970]{dorosh70}
     Doroshkevich A. G., 1970, Afz, 6, 581

\bibitem[\protect\citename{Eisenstein \& Hu }1997]{eishu99}
     Eisenstein D. J., \& Hu W., 1999, ApJ, 511, 5

\bibitem[\protect\citename{Gunn }1967]{gun67} Gunn J. E., 1967, ApJ, 150, 737


\bibitem[\protect\citename{Heavens et al. }2000]{HeaRefHey00}
     Heavens A., Refregier A., Heymans C., 2000, MNRAS, 319, 649 

\bibitem[\protect\citename{Hoyle }1949]{Hoy49}
     Hoyle F., 1949, in Burgers J. M., van de
     Hulst H. C., eds., in  Problems of Cosmical
     Aerodynamics (Dayton, Ohio: Central Air Documents), p. 195

\bibitem[\protect\citename{Hu \& White }2000]{HuWhi00} 
     Hu W. \& White M., 2001, ApJ, 554, 67

\bibitem[\protect\citename{Hu \& White }1997]{HuWhi97} Hu W. \& White M., 1997,
     PhRvD, 56, 596

\bibitem[\protect\citename{Kaiser }1992]{kai92}
     Kaiser N., 1992, ApJ, 388, 272

\bibitem[\protect\citename{Kaiser, Wilson \& Luppino }2000]{KaiWilLup00}
     Kaiser N., Wilson G., Luppino G. A., 2000, [astro-ph/0003338]

\bibitem[\protect\citename{Kamionkowski, Kosowsky \& Stebbins }1997]
     {KamKosSte97} 
     Kamionkowski M., Kosowsky A., Stebbins A., 1997, PhRvD, 55, 7368


\bibitem[\protect\citename{Kamionkowski et al. }1998]{Kametal98}
     Kamionkowski M., Babul A, Cress C., Refregier A., 1998, MNRAS, 301, 1064

\bibitem[\protect\citename{Lee \& Pen }2000]{LeePen00a} 
     Lee J., \& Pen U.-L., 2000, ApJ, 532, L5

\bibitem[\protect\citename{Lee \& Pen }2001]{LeePen00b} 
     Lee J., \& Pen U.-L., 2001, ApJ, 555, 106 

\bibitem[\protect\citename{Maoli et al. }2001]{Maoetal01} 
     Maoli R. et al., 2001, A\&A, 368, 766

\bibitem[\protect\citename{Mellier }1999]{mellier99}
     Mellier Y., 1999, {\em Ann. Rev. Astron. \& Astrophys.}, 37, 127

\bibitem[\protect\citename{Miralda-Escud\'e }1991]{mir91}
     Miralda-Escud\'e J., 1991, ApJ, 380, 1

\bibitem[\protect\citename{Pen, Lee \& Seljak }2000]{penleesel00} 
     Pen U.-L., Lee J., Seljak U., 2000, ApJ, 543 ,L107

\bibitem[\protect\citename{Porciani, Dekel \& Hoffman }2002a]{PorDekHof01a} 
     Porciani C., Dekel A., Hoffman Y., 2002a, MNRAS, 332, 325 

\bibitem[\protect\citename{Porciani, Dekel \& Hoffman }2002b]{PorDekHof01b} 
     Porciani C., Dekel A., Hoffman Y., 2002b, MNRAS, 332, 339 

\bibitem[\protect\citename{Rhodes, Refregier \& Groth }2001]{RhoRefGro01}
     Rhodes J., Refregier A., Groth E. J., 2001, ApJ, 552, L85

\bibitem[\protect\citename{Stebbins }1996]{Ste96}
     Stebbins A., 1996, [astro-ph/9609149]

\bibitem[\protect\citename{Sugerman, Summers \& Kamionkowski }2000]{SugSumKam00}
     Sugerman B., Summers F. J., Kamionkowski M., 2000, MNRAS,
     311, 762

\bibitem[\protect\citename{van Waerbeke et al. }2000]{Waeetal00}
     van Waerbeke L. et al., 2000, A\&A, 358, 30

\bibitem[\protect\citename{Vitvitska et al. }2001]{Vitetal01}
     Vitvitska M. et al, 2001, [astro-ph/0105349]

\bibitem[\protect\citename{White \& Hu }2000]{WhiHu00}
     White M. \& Hu W., 2000, ApJ, 537, 1

\bibitem[\protect\citename{Wittman et al. }2000]{Witetal00}
     Witmann D. A. et al.,
     2000, Nature, 405, 143

\bibitem[\protect\citename{White }1984]{white84}
     White S. D. M., 1984, ApJ, 286, 38

\bibitem[\protect\citename{Zaldarriaga \& Seljak }1997]{ZalSel97} 
     Zaldarriaga M. \& Seljak U., 1997, PhRvD, 55, 1830

\end{thebibliography}
\end{document}